\documentclass[prd,twocolumn,showkeys,superscriptaddress,floatfix]{revtex4}
\usepackage[latin1]{inputenc} 
\usepackage{fontenc}      
\usepackage{graphicx,epsfig,units,color}
\usepackage{amsmath,amssymb}

\newcommand{\bean}{\begin{eqnarray}}
\newcommand{\eean}{\end{eqnarray}}
\newcommand{\be}{\begin{equation}}
\newcommand{\ee}{\end{equation}}

\begin{document}

\title{The supernova Hubble diagram for off-center observers in a spherically symmetric inhomogeneous universe}
\date{\today}

\author{H\aa{}vard Alnes}
\email{havard.alnes@fys.uio.no}
\affiliation{Department of Physics, University of Oslo, PO Box 1048
  Blindern, 0316 Oslo, Norway} 

\author{Morad Amarzguioui}
\email{morad@astro.uio.no}
\affiliation{Institute of Theoretical Astrophysics, University of
  Oslo, PO Box 1029 Blindern, 0315 Oslo, Norway}

\begin{abstract}
We have previously shown that spherically symmetric, inhomogeneous universe models can explain both the supernova data and the location of the first peak in the CMB spectrum without resorting to dark energy. In this work, we investigate whether it is possible to get an even better fit to the supernova data by allowing the observer to be positioned away from the origin in the spherically symmetric coordinate system. In such a scenario, the observer sees an anisotropic relation between redshifts and the luminosity distances of supernovae. The level of anisotropy allowed by the data will then constrain how far away from the origin the observer can be located, and possibly even allow for a better fit. Our analysis shows that the fit is indeed improved, but not by a significant amount. Furthermore, it shows that the supernova data do not place a rigorous constraint on how far off-center the observer can be located. 
\end{abstract}

\maketitle

\section{Introduction}

When analyzed within the framework of homogeneous FRW models, observations indicate that the universe is in a state of accelerated expansion. This is interpreted as proof that the energy content of the universe is  presently dominated by some kind of dark energy with negative pressure \cite{riess98,perlmutter99,tonry03,knop03,Riess:2004nr,tegmark03,Spergel:2003cb,Spergel:2006hy}.
However, for inhomogeneous universe models this needs not be the case. A first step towards gaining an insight into how inhomogeneities affect the observables was made over two decades ago by Partovi and Mashhoon in \cite{Partovi:1982cg} and \cite{Mashhoon:1984}. More recently, the interest in such models has grown increasingly as an alternative way to explain the appararent acceleration without introducing dark energy. A non-exhaustive list of references investigating this possibility is \cite{Pascual-Sanchez:1999zr,celerier:99,Dabrowski:1999sm,Iguchi:2002,alnes:2005,alnes:2006,Alnes:2005nq,Bolejko:2005fp,Mansouri:2005rf,
Vanderveld:2006rb,Rakic:2006tp,Garfinkle:2006sb,Biswas:2006ub,Rasanen:2006kp,Enqvist:2006cg,Celerier:2006gy,
Chung:2006xh}.

%

In \cite{alnes:2005} we showed that an inhomogeneous, but spherically symmetric, matter-dominated universe model can easily explain the supernova Hubble diagram. Due to lack of a complete understanding of how perturbations behave in such inhomogeneous models, it is at present time not possible to calculate a theoretical CMB power spectrum to compare with that measured by the WMAP satellite \cite{Hinshaw:2006ia}. However, one can still extract some information from the CMB, in the form of the location of the first peak in the power spectrum, that can be used to constrain the model. In \cite{alnes:2005}, it is shown that specific spherically symmetric models can account for both the supernova Hubble diagram and the location of the first CMB peak. 

In such a scenario, the "accelerated expansion" is not a real effect, but is explained by the fact that the local expansion rate no longer is the same everywhere in space at a given time. Since the observations probe the expansion rate along the past light-cone, which at different redshifts corresponds to different points in space, the observed increasing expansion rate can just as well be explained by an appropriate spatially varying expansion rate. Qualitatively, the expansion rate would have to increase radially towards the observer.
A specific model was found in Ref.~\cite{alnes:2005} where there is a very good agreement with both SNIa data and the location of the first CMB peak. In this model, the observer was located at the center of a large underdensity in an otherwise flat and matter dominated universe. Although the model is able to explain these data without the introduction of dark energy, it does so at the expense of the Copernican Principle, since the observer is located at a unique place in space. A more philosophically appealing scenario would be one where the observer does not necessarily have to be positioned exactly at the center of the underdensity.  

For an off-center observer in our model, the universe appears to be anisotropic. In an earlier work \cite{alnes:2006}, we showed that this anisotropy induces additional anisotropies in the temperature map on top of the intrinsic anisotropies from the CMB. The farther away from the origin the observer is, the greater these induced anisotropies are. Thus, the observed CMB imposes a constraint on how far away from the origin the observer can be located. In \cite{alnes:2006} we found the principal constraint to arise from the dipole, with a largest possible displacement from the origin of around $\unit[15]{Mpc}$. Similarly, the luminosity distance-redshift relationship will also become anisotropic. This might provide further constrains on the position of the observer inside the underdensity from an independent data set, namely the supernova observations. 

In addition, it is in principle an interesting question whether the supernova data lend support to the idea of anisotropic surroundings, since the CMB
indicates that the universe is extremely isotropic on large scales. This has previously been studied by Kolatt and Lahav \cite{Kolatt:2000} and Bochner \cite{Bochner:2004}. Although their approaches were somewhat different, they both looked in essence for evidence of a lack of uniformity in the Hubble flow from a statistical point of view. They both conclude that there is no statistically significant evidence for discarding the hypothesis of a homogeneous universe. However, the advantage of our approach is that we can compare the data with a specific model, with actual predictions for the luminosity distance-redshift relation in different directions, rather than  just treating angular variations in the supernova data with statistical tools. Therefore, our method could yield interesting results despite the low number of supernova observations available. 

The structure of this paper is as follows. In Sect.~\ref{sec:geodesics} we present the solution to the field equations for the model in \cite{alnes:2005} and look specifically at how the luminosity distance-redshift relation is altered for an off-center observer. In Sect.~\ref{sec:results} we calculate the theoretical luminosity distance-redshift relation and compare it with that inferred from the supernovae in the Riess et al. Gold Set. Finally, in Sect.~\ref{sec:disc} we summarize our work and discuss the implications of our results.

\section{The luminosity distance in an inhomogeneous universe}
\label{sec:geodesics}
The universe model we will focus on in this work is that described in \cite{alnes:2005}. This is a spherically symmetric space-time, which is described by the Lema\^{\i}tre-Tolman-Bondi (LTB) metric \cite{Lemaitre:1933,Tolman:1934,Bondi:1947}. The line element can be written as
\be
\label{eq:metric}
ds^2 = -dt^2 + \frac{[R'(r,t)]^2}{1+\beta(r)}dr^2+R^2(r,t)\left[d\theta^2+\sin^2\theta d\phi^2\right], 
\ee
where $R(r,t)$ is a position-dependent scale factor, and $\beta(r)$ is
related to the spatial curvature. For an extensive review of the properties of this model the interested reader is referred to Ref.~\cite{Krasinski:1997}.
 
For LTB models that contain only matter, the Einstein equations can be solved analytically. The exact form of the solutions depends on the sign of $\beta(r)$. For a model with a positive $\beta(r)$, corresponding to negative local spatial curvature, the solution can be written parametrically in terms of a conformal time $\eta$
\bean
R &=& \frac{\alpha}{2\beta}(\cosh \eta - 1)\nonumber\\
  &&+R_0\left[\cosh \eta +
  \sqrt{\frac{\alpha+\beta R_0}{\beta R_0}}\sinh \eta\right],\\ 
\sqrt{\beta}t &=& \frac{\alpha}{2\beta}(\sinh \eta-
\eta)\nonumber\\&&
+R_0\left[\sinh \eta + \sqrt{\frac{\alpha+\beta R_0}{\beta 
R_0}}\left(\cosh \eta   -1\right)\right], 
\eean
where $\alpha(r)$ is related to the density of matter. We must assume $\alpha(r)>0$ for the matter density to remain positive everywhere. Furthermore, we have defined $R_0 \equiv R(r,0)$, interpreting $t=0$ as the time of last scattering, when photons decoupled from baryons.

The scenario considered in \cite{alnes:2005} was one where the observer is at the center of an underdense bubble. The angular diameter distance is then  isotropic and is given by
\be
\label{eq:angdisthomo}
d_A(z) = R(\hat{r})\,.
\ee
where $\hat{r}(z)$ describes the past light-cone of the observer at $t=t_0$. As indicated in the introduction, if the observer is placed at an off-center location, the distance measures will become anisotropic. The explicit effect this has on the expression for the angular diameter distance has been analyzed previously, first by Ellis et al.~\cite{ellis:1985}, and also later by Humphreys et al.~\cite{Humphreys:1996fd} and Biswas et al.~\cite{Biswas:2006ub}.
We use the expression presented in \cite{Humphreys:1996fd}, which reads
\be
\label{eq:d_A}
d_A^4 \sin^2\!\gamma = \tilde{g}_{\gamma\gamma}\tilde{g}_{\xi\xi}-\tilde{g}_{\gamma\xi}^2\,.
\ee
where $\tilde{g}_{\mu\nu}$ is the observer frame, i.e. a frame centered on the observed in which the angular part of the line element is
\be
d\Omega^2 = d\gamma^2+\sin^2\!\gamma d\xi^2\,.
\ee
The angles $\gamma$ and $\xi$ correspond to the polar and azimuth angles in this frame.

In our case, such a frame can be constructed using the light-cones specified by ($t$,$t_0$,$\gamma$,$\xi$), where $t$ is cosmic time, and $t_0$ is the time when the photons hit the observer at angles $\gamma$ and $\xi$. In order to simplify the calculations, but still keep the results completely general, we choose the $z$-axis in the direction of the off-center observer. The spatial coordinates of the observer in the reference frame defined by the metric Eq.~\eqref{eq:metric} are then $r=r_{\text{obs}}$ and $\theta=0$, with $\phi$ being degenerate. Transforming the coordinates for the photon trajectories back to this frame from the observer frame, we get 
\bean
t &=& t\\
r &=& \hat{r}(t,t_0,\gamma)\\
\theta &=& \hat{\theta}(t,t_0,\gamma)\\
\phi &=& \xi
\eean
where the $\hat{r}$ and $\hat{\theta}$ functions are solutions to the geodesic equation as a function of $t$ and the initial conditions $r_0$, $t_0$ and $\gamma$. Note that due to the axial symmetry about the $z$-axis, these functions do not depend on the azimuth angle $\xi$. A more thorough discussion of these solutions and their implications can be found in Ref.~\cite{alnes:2006}. 

The metric $\tilde{g}_{\mu\nu}$ in the observers local coordinate system is given by a simple coordinate transformation, i.e.
\be
\tilde{g}_{\mu\nu} = g_{\sigma\tau}\frac{\partial x^\sigma}{\partial \tilde{x}^\mu}\frac{\partial x^\tau}{\partial \tilde{x}^\nu}\,,
\ee
which yields the following components
\bean
\tilde{g}_{\gamma\gamma} &=& g_{rr}\left(\frac{\partial \hat{r}}{\partial \gamma}\right)^2+g_{\theta\theta}\left(\frac{\partial \hat{\theta}}{\partial
  \gamma}\right)^2,\\
\tilde{g}_{\xi\xi} &=& g_{\phi\phi} = R^2\sin^2\theta\,,\\
\tilde{g}_{\gamma\xi} &=& 0\,.
\eean
Substituting these into Eq.~\eqref{eq:d_A}, we arrive at the following expression for the angular diameter distance 
\be
\label{eq:angdiam}
d_A^4 = \frac{R^4 \sin^2\theta}{\sin^2\gamma} \left[\frac{(R')^2}{R^2(1+\beta)}\left(\frac{\partial \hat{r}}{\partial \gamma}\right)^2+\left(\frac{\partial
    \hat{\theta}}{\partial \gamma}\right)^2\right]\,.
\ee
As a simple consistency check, we should recover the usual expression in Eq.~\eqref{eq:angdisthomo} in the limit where the observer is at the origin of the coordinate system. In this limit, the two angles $\theta$ and $\gamma$ coincide and the radial coordinate becomes independent of $\gamma$. This means that the two partial derivatives in Eq.~(\ref{eq:angdiam}) become 0 and 1, respectively, and we end up with Eq.~(\ref{eq:angdisthomo}) as expected.

The distance measure probed by the supernova data is not the angular diameter distance, but rather the luminosity distance $d_L$. However, there exists a general relation \cite{Ellis71} which allows us to relate these two distance measures:
\be \label{eq:dl}
d_L = (1+z)^2 d_A\,.
\ee
Roughly, the two factors of $(1+z)$ take into account the reduction of energy per photon due to redshifting and the reduced arrival rate of incoming photons due to time dilation.

Replicating the approach in \cite{alnes:2006}, we solve the geodesic equations which determine the path of infalling photons relative to an off-center observer. Next, we use Eqs.~\eqref{eq:angdiam} and \eqref{eq:dl} to obtain a theoretical prediction for the luminosity distance-redshift relation in this scenario. However, what we measure when we observe the supernovae is not the luminosity distance directly, but rather the apparent magnitude. These can be related to each other through the expression
\be \label{eq:distmod}
	\mu = 5\log_{10}\!\frac{d_L}{\unit[]{Mpc}}+25\,,
\ee
where $\mu$ is the distance modulus, which is simply the apparent magnitude minus an absolute magnitude \cite{Weinberg:1972}.

\section{Results}
\label{sec:results}
In the scenario which we considered in \cite{alnes:2005}, the functions $\alpha(r)$ and $\beta(r)$ were parametrized as
\bean
\alpha(r) &=& H_{0}^2r^3\left[ \alpha_0-\Delta \alpha
  \left(\frac{1}{2}-\frac{1}{2}\tanh \frac{r-r_0}{2\Delta r}\right)
  \right]\\  
\beta(r) &=& H_{0}^2r^2\left[ \beta_0 - \Delta \beta 
  \left(\frac{1}{2}-\frac{1}{2}\tanh \frac{r-r_0}{2\Delta r}\right)
  \right]
\eean
This corresponds to a smooth interpolation between two homogeneous regions where the inner region has a lower matter density than the outer region, thus describing a spherical bubble in an otherwise homogeneous universe. The parameter $H_{0} = 100\, h_{\text{out}} \unit{km\, s^{-1} Mpc^{-1}}$ is the Hubble constant of the outer homogeneous region today, while $\alpha_0$ and $\beta_0$ are the relative
densities of matter and curvature in this region. Furthermore, $\Delta\alpha$ and $\Delta\beta$ determine the difference in matter density and curvature between the regions, while $r_0$ and $\Delta r$ specify the position and width of the transition. In the original analysis we restricted the parameter space by imposing the constraints $\beta_0=1-\alpha_0$ and $\Delta\beta=-\Delta\alpha$. We will keep these constraints in the present analysis too.

\begin{figure}
\begin{center}
\includegraphics[width=8cm]{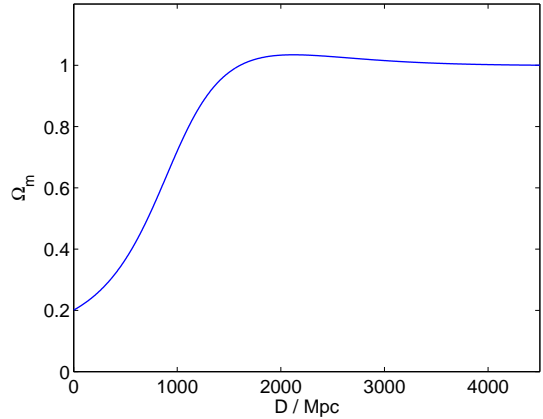}
\end{center}
\caption{The relative matter density today as a function of physical distance
  from the center.}
\label{fig:O_m}
\end{figure}

Assuming that the observer was positioned at the center of the bubble, we then found a model that gave a good agreement with the Hubble diagram of observed SNIa and the position of the first CMB peak. The properties of this model are listed in table~\ref{tab:model}, and the corresponding density profile of the underdensity is plotted in figure \ref{fig:O_m}. The generalized matter density used in this plot is defined in \cite{alnes:2005}.

\squeezetable
\begin{table}[t]
\begin{tabular}{lcc}
\hline 
Description & Symbol & Value \\
\hline 
Density contrast parameter & $\Delta \alpha$         & 0.90   \\
Physical distance out to transition point [Gpc] & $D_0$                        & 1.34   \\
Transition width & $\Delta r/r_0$                    & 0.40   \\
Fit to supernovae & $\chi^2_{SN}$                    & 176.2 \\
Position of first CMB peak & $\mathcal{S}$           & 1.006 \\
Age of the universe [Gyr] & $t_0$                    & 12.8  \\
Relative density inside underdensity & $\Omega_{m,in}$ & 0.20  \\
Relative density outside underdensity & $\Omega_{m,out}$ & 1.00 \\
Hubble parameter inside underdensity & $h_{in}$ & 0.65  \\
Hubble parameter outside underdensity & $h_{out}$ & 0.51 \\
\hline
\end{tabular}
\caption{The parameters and properties of the inhomogeneous model which was studied in \cite{alnes:2005}, where the observer is located at the center of the inhomogeneity.}
\label{tab:model}
\end{table}

\begin{figure}
\begin{center}
\includegraphics[width=8.0cm]{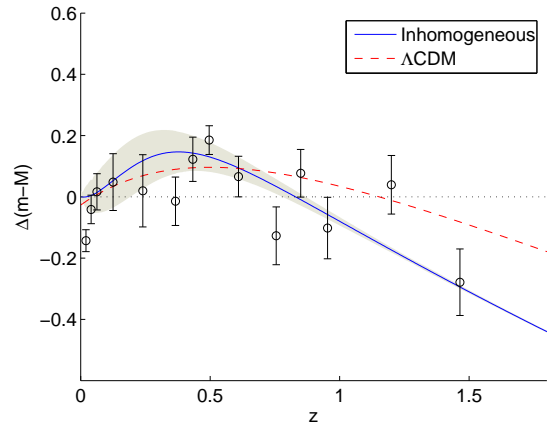}
\end{center}
\caption{The average distance modulus for an off-center observer in our model. The shaded area represents the RMS deviation from the average across the sky. The data points and error bars are binned data from the Riess et al. Gold Set, while the red dashed line is the corresponding best-fit $\Lambda$CDM model.} 
\label{fig:hubble_diagram}
\end{figure}

As explained in the preceding sections, we will investigate what effect moving the observer away from the center of the inhomogeneity has upon the luminosity distance-redshift relation. To get an idea of how much the luminosity distance then varies across the sky, we have plotted in Fig.~\ref{fig:hubble_diagram} the distance modulus as seen by an observer located at a physical distance $D_{obs} = \unit[200]{Mpc}$ from the center. The blue line in the plot represents the average distance modulus, while the shaded area is the RMS deviation from the mean. Also plotted in the same figure are binned data points from the Riess et al. Gold Sample (RGS) of supernovae \cite{Riess:2004nr} and the best-fit $\Lambda$CDM model.  It is evident that our model allow for relatively large variations in the distance modulus for redshifts $z<1$ for off-center observers. One can therefore get a potentially even better fit to the data by placing the observer away from the origin, and taking into account the directions in which the various supernovae have been observed. 

\begin{figure}
\begin{center}
\includegraphics[width=9.0cm]{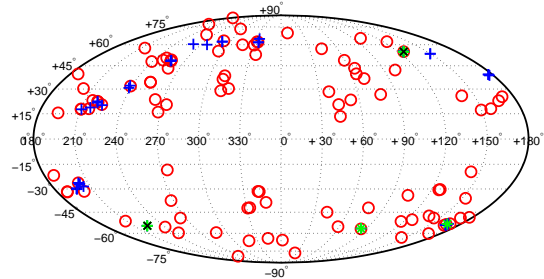}
\end{center}
\caption{The directions on the sky for $157$ supernovae of the Riess et al. Gold Sample. Supernovae with $z<0.5$ are marked with red circles, while supernovae with $0.5<z<1.0$ are marked with blue pluses, $1.0<z<1.5$ with green stars and $z>1.5$ with black crosses.} 
\label{fig:riess_sphere}
\end{figure}

The coordinates of all observed supernovae can be found at the CBAT website \footnote{http://cfa-www.harvard.edu/iau/lists/Supernovae.html}. The angular coordinates of the supernovae in the RGS are plotted in Fig.~\ref{fig:riess_sphere}. The marker and color coding in the figure is as follows. Red circles correspond to supernovae with $z<0.5$, blue pluses to $0.5<z<1.0$, green stars to $1.0<z<1.5$ and black crosses to $z>1.5$. As we can see, the supernovae appear to be distributed relatively evenly across the sky except for in the proximity of the galactic plane.  

%

\begin{figure}
\begin{center}
\includegraphics[width=8.0cm]{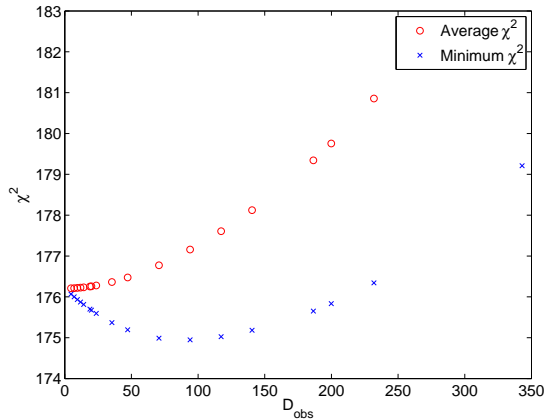}
\end{center}
\caption{The $\chi^2$ fit to observed supernovae as a function of the observer's position. The red circles show the angle-averaged values, while the blue crosses show the angle-optimized values.}
\label{fig:chi2values}
\end{figure}

By moving the observer away from the center of the inhomogeneity, we add three additional degrees of freedom to the original model. These are the radial displacement from the center and the two angles that specify the direction of the displacement. We can treat these extra degrees of freedom in two different ways according to how we deal with the angles. First, we can average over all possible orientations of the underdensity. Basically, what we do then is throw away all information about angular dependence. If we didn't have any information about the direction in which the supernovae occurred, this would be the appropriate way to calculate the fit. The resulting $\chi^2$ value as a function of the observer's position is plotted as red circles in Fig.~\ref{fig:chi2values}. It is clear that the fit becomes increasingly worse the farther away from the center the observer is placed. Thus, if one takes into account only the redshift and neglect information about the exact direct direction on the sky of the supernovae, the data disfavor an off-center observer.   

The other approach is to make use of the explicit angular dependence of the observed supernovae. Since this information is actually available to us, this would seem to be the more fitting way to do the analysis. The approach then is to minimize the $\chi^2$ with respect to the two angles for each radius. These minimal $\chi^2$ values are plotted as blue crosses in Fig.~\ref{fig:chi2values}. It is evident from this plot that one can indeed reduce the residuals by moving the observer away from the center, if one also takes into account the specific directions in the sky of the supernovae.

\begin{figure}
\begin{center}
\includegraphics[width=9.0cm]{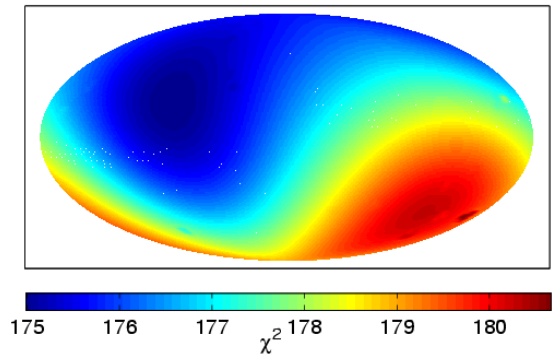}
\end{center}
\caption{The $\chi^2$ residual for supernovae seen by an observer at a radial distance of $D_{obs} = \unit[94]{Mpc}$. The plot shows the residuals as a function of the direction towards the center of the underdensity.} 
\label{fig:skymap}
\end{figure}

The minimized $\chi^2$ value is smallest for an observer located at a physical distance of around $D_{obs} = \unit[94]{Mpc}$ from the center of the inhomogeneity. In Fig.~\ref{fig:skymap}, we have plotted the $\chi^2$ seen by an observer at this distance as a function of the direction towards the center of inhomogeneity . The $\chi^2$ reaches a minimum at $\chi^2_{min} = 174.9$, with a direction towards the center of the inhomogeneity of $(l, b) = (271^\circ, 21^\circ)$.

As we discuss in \cite{alnes:2006}, such an off-center placement of the observer will necessarily induce an additional dipole in the temperature of the cosmic microwave background. CMB photons arriving at the observer's from the direction of the center of the inhomogeneity will have traveled through a larger region with a high Hubble parameter compared to those from the opposite direction. They will therefore be more redshifted, and hence, appear to have lower temperatures. The COBE satellite \cite{Bennett:1996ce} shows that the measured dipole in the background temperature points in the direction $(l, b) = (264^\circ, 48^\circ)$. This means that the temperature is perceived to be higher in this direction compared to the opposite direction. Thus it appears that the measured and the induced dipole for the best-fit off-center observer point in almost the opposite directions. However, this does not necessarily represent a problem, since the effect can be countered by an appropriate peculiar velocity of our galaxy. We should therefore not expect a correlation between the direction of the measured dipole and the direction towards the center of the inhomogeneity. 

\section{Discussion}
\label{sec:disc}
In this work we investigated the inhomogeneous but spherically symmetric model which we have explored previously in \cite{alnes:2005} and \cite{alnes:2006}. In this model the observer is located inside an underdensity in an otherwise homogeneous and flat universe, described by the Lema\^{\i}tre-Tolman-Bondi space-time. In \cite{alnes:2005} we showed that it is possible for such models to explain both the supernova data and the position of the first peak in the CMB power spectrum. The aim of the current work was to learn whether it is possible to improve the fit to the supernova data by moving the observer away from the center of the inhomogeneity, and also how much these data constrain such a movement. In addition to possibly providing a better fit to the data, allowing the observer to be located at a different position than the dead center of the inhomogeneity is also more appealing from a purely philosophical perspective. After all, if we imagine placing several imaginary observer randomly inside the underdensity, the chance of any of them ending up at the exact center vanishes. On the other hand, there is a finite probability that an observer will end up inside any region with a finite radius.

The universe no longer appears to be isotropic for such an off-center observer. This induced anisotropy is an effect that should be detectable in the observations. Specifically for the supernova observations, this manifests itself in the form of an anisotropic relation between the luminosity distance and the redshift. For such an observer, supernovae at the same redshifts but at different direction in the sky, do not necessarily shine with the same brightness.

We placed the observer at different radii from the center of the underdensity, and for each radius chose the direction from the observer to the center to be that which optimized the fit. In this way we were able to improve the fit compared to the isotropic observer. The fit was found to be optimal for a radius of $D_{obs} = \unit[94]{Mpc}$ in the direction $(l, b) = (271^\circ,21^\circ)$ in galactic coordinates. However, the improvement turns out to be only slight. The minimal $\chi^2$ for an off-center observer is $174.9$ compared to $176.2$ for an observer at the center. Although the $\chi^2$ has been reduced, the fit can not be said to have improved, considering that the off-center placement adds additional degrees of freedom. In fact, the $\chi^2$ per degrees of freedom is higher for all off-center observers compared to observers at the center.   

Looking at Fig.~\ref{fig:chi2values}, we see that the $\chi^2$ is lower for off-center observers for radial distances out to about $D=\unit[225]{Mpc}$. From this we can conclude that anisotropies in the supernova data do not constrain very well how far away from the center the observer can be located. In \cite{alnes:2006} we looked at such constraints arising from the dipole in the CMB temperature. We found that the observer had to remain within a radius of around $\unit[15]{Mpc}$ for the dipole to be be in agreement with COBE measurements. It is clear that the current supernova data cannot improve this constraint any further.

The main conclusion we can draw from our analysis is that the current supernova data do not offer any substantial evidence for an off-center observer. Nor does it constrain very well how far off-center such observers can be located. This is partly due to the fact that there are too few supernovae in the sample. In the future, substantially larger and better supernova samples will be available, such as those which will be provided by the Supernova Legacy Survey \footnote{http://www.cfht.hawaii.edu/SNLS/}, and hopefully the Supernova Acceleration Probe (SNAP) \footnote{http://snap.lbl.gov/}. These might allow us to draw stronger conclusions regarding the anisotropy of the local universe. 

\begin{acknowledgments}
The authors wish to thank Teppo Mattsson for helpful suggestions and critique. MA acknowledges support from the Norwegian Research Council through the project "Shedding Light on Dark Energy" (grant 159637/V30).
\end{acknowledgments}

\bibliography{main.bbl}

\begin{thebibliography}{39}
\expandafter\ifx\csname natexlab\endcsname\relax\def\natexlab#1{#1}\fi
\expandafter\ifx\csname bibnamefont\endcsname\relax
  \def\bibnamefont#1{#1}\fi
\expandafter\ifx\csname bibfnamefont\endcsname\relax
  \def\bibfnamefont#1{#1}\fi
\expandafter\ifx\csname citenamefont\endcsname\relax
  \def\citenamefont#1{#1}\fi
\expandafter\ifx\csname url\endcsname\relax
  \def\url#1{\texttt{#1}}\fi
\expandafter\ifx\csname urlprefix\endcsname\relax\def\urlprefix{URL }\fi
\providecommand{\bibinfo}[2]{#2}
\providecommand{\eprint}[2][]{\url{#2}}

\bibitem[{\citenamefont{Riess et~al.}(1998)}]{riess98}
\bibinfo{author}{\bibfnamefont{A.~G.} \bibnamefont{Riess}}
  \bibnamefont{et~al.}, \bibinfo{journal}{Astron. J.}
  \textbf{\bibinfo{volume}{116}}, \bibinfo{pages}{1009} (\bibinfo{year}{1998}),
  \eprint{astro-ph/9805201}.

\bibitem[{\citenamefont{Perlmutter et~al.}(1999)}]{perlmutter99}
\bibinfo{author}{\bibfnamefont{S.}~\bibnamefont{Perlmutter}}
  \bibnamefont{et~al.}, \bibinfo{journal}{Astrophys. J.}
  \textbf{\bibinfo{volume}{517}}, \bibinfo{pages}{565} (\bibinfo{year}{1999}),
  \eprint{astro-ph/9812133}.

\bibitem[{\citenamefont{Tonry et~al.}(2003)}]{tonry03}
\bibinfo{author}{\bibfnamefont{J.~L.} \bibnamefont{Tonry}}
  \bibnamefont{et~al.}, \bibinfo{journal}{Astrophys. J.}
  \textbf{\bibinfo{volume}{594}}, \bibinfo{pages}{1} (\bibinfo{year}{2003}),
  \eprint{astro-ph/0305008}.

\bibitem[{\citenamefont{Knop et~al.}(2003)}]{knop03}
\bibinfo{author}{\bibfnamefont{R.~A.} \bibnamefont{Knop}} \bibnamefont{et~al.},
  \bibinfo{journal}{Astrophys. J.} \textbf{\bibinfo{volume}{598}},
  \bibinfo{pages}{102} (\bibinfo{year}{2003}), \eprint{astro-ph/0309368}.

\bibitem[{\citenamefont{Riess et~al.}(2004)}]{Riess:2004nr}
\bibinfo{author}{\bibfnamefont{A.~G.} \bibnamefont{Riess}} \bibnamefont{et~al.}
  (\bibinfo{collaboration}{Supernova Search Team}),
  \bibinfo{journal}{Astrophys. J.} \textbf{\bibinfo{volume}{607}},
  \bibinfo{pages}{665} (\bibinfo{year}{2004}), \eprint{astro-ph/0402512}.

\bibitem[{\citenamefont{Tegmark et~al.}(2004)}]{tegmark03}
\bibinfo{author}{\bibfnamefont{M.}~\bibnamefont{Tegmark}} \bibnamefont{et~al.}
  (\bibinfo{collaboration}{SDSS collaboration}), \bibinfo{journal}{Phys. Rev.}
  \textbf{\bibinfo{volume}{D69}}, \bibinfo{pages}{103501}
  (\bibinfo{year}{2004}), \eprint{astro-ph/0310723}.

\bibitem[{\citenamefont{Spergel et~al.}(2003)}]{Spergel:2003cb}
\bibinfo{author}{\bibfnamefont{D.~N.} \bibnamefont{Spergel}}
  \bibnamefont{et~al.} (\bibinfo{collaboration}{WMAP}),
  \bibinfo{journal}{Astrophys. J. Suppl.} \textbf{\bibinfo{volume}{148}},
  \bibinfo{pages}{175} (\bibinfo{year}{2003}), \eprint{astro-ph/0302209}.

\bibitem[{\citenamefont{Spergel et~al.}(2006)}]{Spergel:2006hy}
\bibinfo{author}{\bibfnamefont{D.~N.} \bibnamefont{Spergel}}
  \bibnamefont{et~al.} (\bibinfo{year}{2006}), \eprint{astro-ph/0603449}.

\bibitem[{\citenamefont{Partovi and Mashhoon}(1984)}]{Partovi:1982cg}
\bibinfo{author}{\bibfnamefont{M.~H.} \bibnamefont{Partovi}} \bibnamefont{and}
  \bibinfo{author}{\bibfnamefont{B.}~\bibnamefont{Mashhoon}},
  \bibinfo{journal}{Astrophys. J.} \textbf{\bibinfo{volume}{276}},
  \bibinfo{pages}{4} (\bibinfo{year}{1984}).

\bibitem[{\citenamefont{{Mashhoon}}(1984)}]{Mashhoon:1984}
\bibinfo{author}{\bibfnamefont{B.}~\bibnamefont{{Mashhoon}}}, in
  \emph{\bibinfo{booktitle}{Big-Bang Cosmology Symposium in honour of G.
  Lema\^{\i}tre}}, edited by
  \bibinfo{editor}{\bibfnamefont{A.}~\bibnamefont{{Berger}}}
  (\bibinfo{year}{1984}), pp. \bibinfo{pages}{75--81}.

\bibitem[{\citenamefont{Pascual-Sanchez}(1999)}]{Pascual-Sanchez:1999zr}
\bibinfo{author}{\bibfnamefont{J.~F.} \bibnamefont{Pascual-Sanchez}},
  \bibinfo{journal}{Mod. Phys. Lett.} \textbf{\bibinfo{volume}{A14}},
  \bibinfo{pages}{1539} (\bibinfo{year}{1999}), \eprint{gr-qc/9905063}.

\bibitem[{\citenamefont{C\'{e}l\'{e}rier}(2000)}]{celerier:99}
\bibinfo{author}{\bibfnamefont{M.-N.} \bibnamefont{C\'{e}l\'{e}rier}},
  \bibinfo{journal}{Astron. Astrophys.} \textbf{\bibinfo{volume}{353}},
  \bibinfo{pages}{63} (\bibinfo{year}{2000}), \eprint{astro-ph/9907206}.

\bibitem[{\citenamefont{Dabrowski}(2002)}]{Dabrowski:1999sm}
\bibinfo{author}{\bibfnamefont{M.~P.} \bibnamefont{Dabrowski}},
  \bibinfo{journal}{Grav. Cosmol.} \textbf{\bibinfo{volume}{8}},
  \bibinfo{pages}{190} (\bibinfo{year}{2002}), \eprint{gr-qc/9905083}.

\bibitem[{\citenamefont{Iguchi et~al.}(2002)\citenamefont{Iguchi, Nakamura, and
  Nakao}}]{Iguchi:2002}
\bibinfo{author}{\bibfnamefont{H.}~\bibnamefont{Iguchi}},
  \bibinfo{author}{\bibfnamefont{T.}~\bibnamefont{Nakamura}}, \bibnamefont{and}
  \bibinfo{author}{\bibfnamefont{K.}~\bibnamefont{Nakao}},
  \bibinfo{journal}{Prog. Theor. Phys.} \textbf{\bibinfo{volume}{108}},
  \bibinfo{pages}{809} (\bibinfo{year}{2002}).

\bibitem[{\citenamefont{Alnes et~al.}(2006)\citenamefont{Alnes, Amarzguioui,
  and Gr\o{}n}}]{alnes:2005}
\bibinfo{author}{\bibfnamefont{H.}~\bibnamefont{Alnes}},
  \bibinfo{author}{\bibfnamefont{M.}~\bibnamefont{Amarzguioui}},
  \bibnamefont{and} \bibinfo{author}{\bibfnamefont{O.}~\bibnamefont{Gr\o{}n}},
  \bibinfo{journal}{Phys. Rev.} \textbf{\bibinfo{volume}{D73}},
  \bibinfo{pages}{083519} (\bibinfo{year}{2006}), \eprint{astro-ph/0512006}.

\bibitem[{\citenamefont{Alnes and Amarzguioui}(2006)}]{alnes:2006}
\bibinfo{author}{\bibfnamefont{H.}~\bibnamefont{Alnes}} \bibnamefont{and}
  \bibinfo{author}{\bibfnamefont{M.}~\bibnamefont{Amarzguioui}}
  (\bibinfo{year}{2006}), \eprint{astro-ph/0607334}.

\bibitem[{\citenamefont{Alnes et~al.}(2005)\citenamefont{Alnes, Amarzguioui,
  and Gron}}]{Alnes:2005nq}
\bibinfo{author}{\bibfnamefont{H.}~\bibnamefont{Alnes}},
  \bibinfo{author}{\bibfnamefont{M.}~\bibnamefont{Amarzguioui}},
  \bibnamefont{and} \bibinfo{author}{\bibfnamefont{O.}~\bibnamefont{Gron}}
  (\bibinfo{year}{2005}), \eprint{astro-ph/0506449}.

\bibitem[{\citenamefont{Bolejko}(2005)}]{Bolejko:2005fp}
\bibinfo{author}{\bibfnamefont{K.}~\bibnamefont{Bolejko}}
  (\bibinfo{year}{2005}), \eprint{astro-ph/0512103}.

\bibitem[{\citenamefont{Mansouri}(2005)}]{Mansouri:2005rf}
\bibinfo{author}{\bibfnamefont{R.}~\bibnamefont{Mansouri}}
  (\bibinfo{year}{2005}), \eprint{astro-ph/0512605}.

\bibitem[{\citenamefont{Vanderveld et~al.}(2006)\citenamefont{Vanderveld,
  Flanagan, and Wasserman}}]{Vanderveld:2006rb}
\bibinfo{author}{\bibfnamefont{R.~A.} \bibnamefont{Vanderveld}},
  \bibinfo{author}{\bibfnamefont{E.~E.} \bibnamefont{Flanagan}},
  \bibnamefont{and}
  \bibinfo{author}{\bibfnamefont{I.}~\bibnamefont{Wasserman}},
  \bibinfo{journal}{Phys. Rev.} \textbf{\bibinfo{volume}{D74}},
  \bibinfo{pages}{023506} (\bibinfo{year}{2006}), \eprint{astro-ph/0602476}.

\bibitem[{\citenamefont{Raki\'c et~al.}(2006)\citenamefont{Raki\'c,
  R\"{a}s\"{a}nen, and Schwarz}}]{Rakic:2006tp}
\bibinfo{author}{\bibfnamefont{A.}~\bibnamefont{Raki\'c}},
  \bibinfo{author}{\bibfnamefont{S.}~\bibnamefont{R\"{a}s\"{a}nen}},
  \bibnamefont{and} \bibinfo{author}{\bibfnamefont{D.~J.}
  \bibnamefont{Schwarz}}, \bibinfo{journal}{Mon. Not. Roy. Astron. Soc. Lett.}
  \textbf{\bibinfo{volume}{369}}, \bibinfo{pages}{L27} (\bibinfo{year}{2006}),
  \eprint{astro-ph/0601445}.

\bibitem[{\citenamefont{Garfinkle}(2006)}]{Garfinkle:2006sb}
\bibinfo{author}{\bibfnamefont{D.}~\bibnamefont{Garfinkle}},
  \bibinfo{journal}{Class. Quant. Grav.} \textbf{\bibinfo{volume}{23}},
  \bibinfo{pages}{4811} (\bibinfo{year}{2006}), \eprint{gr-qc/0605088}.

\bibitem[{\citenamefont{Biswas et~al.}(2006)\citenamefont{Biswas, Mansouri, and
  Notari}}]{Biswas:2006ub}
\bibinfo{author}{\bibfnamefont{T.}~\bibnamefont{Biswas}},
  \bibinfo{author}{\bibfnamefont{R.}~\bibnamefont{Mansouri}}, \bibnamefont{and}
  \bibinfo{author}{\bibfnamefont{A.}~\bibnamefont{Notari}}
  (\bibinfo{year}{2006}), \eprint{astro-ph/0606703}.

\bibitem[{\citenamefont{Räsänen}(2006)}]{Rasanen:2006kp}
\bibinfo{author}{\bibfnamefont{S.}~\bibnamefont{Räsänen}}
  (\bibinfo{year}{2006}), \eprint{astro-ph/0607626}.

\bibitem[{\citenamefont{Enqvist and Mattsson}(2006)}]{Enqvist:2006cg}
\bibinfo{author}{\bibfnamefont{K.}~\bibnamefont{Enqvist}} \bibnamefont{and}
  \bibinfo{author}{\bibfnamefont{T.}~\bibnamefont{Mattsson}}
  (\bibinfo{year}{2006}), \eprint{astro-ph/0609120}.

\bibitem[{\citenamefont{C\'el\'erier}(2006)}]{Celerier:2006gy}
\bibinfo{author}{\bibfnamefont{M.-N.} \bibnamefont{C\'el\'erier}}
  (\bibinfo{year}{2006}), \eprint{astro-ph/0609352}.

\bibitem[{\citenamefont{Chung and Romano}(2006)}]{Chung:2006xh}
\bibinfo{author}{\bibfnamefont{D.~J.~H.} \bibnamefont{Chung}} \bibnamefont{and}
  \bibinfo{author}{\bibfnamefont{A.~E.} \bibnamefont{Romano}}
  (\bibinfo{year}{2006}), \eprint{astro-ph/0608403}.

\bibitem[{\citenamefont{Hinshaw et~al.}(2006)}]{Hinshaw:2006ia}
\bibinfo{author}{\bibfnamefont{G.}~\bibnamefont{Hinshaw}} \bibnamefont{et~al.}
  (\bibinfo{year}{2006}), \eprint{astro-ph/0603451}.

\bibitem[{\citenamefont{Kolatt and Lahav}(2001)}]{Kolatt:2000}
\bibinfo{author}{\bibfnamefont{T.~S.} \bibnamefont{Kolatt}} \bibnamefont{and}
  \bibinfo{author}{\bibfnamefont{O.}~\bibnamefont{Lahav}},
  \bibinfo{journal}{Mon. Not. Roy. Astron. Soc.}
  \textbf{\bibinfo{volume}{323}}, \bibinfo{pages}{859} (\bibinfo{year}{2001}),
  \eprint{astro-ph/0008041}.

\bibitem[{\citenamefont{Bochner}(2004)}]{Bochner:2004}
\bibinfo{author}{\bibfnamefont{B.}~\bibnamefont{Bochner}},
  \bibinfo{journal}{ECONF} \textbf{\bibinfo{volume}{C041213}},
  \bibinfo{pages}{1301} (\bibinfo{year}{2004}).

\bibitem[{\citenamefont{Lema\^itre}(1933)}]{Lemaitre:1933}
\bibinfo{author}{\bibfnamefont{G.}~\bibnamefont{Lema\^itre}},
  \bibinfo{journal}{Annales Soc. Sci. Brux.} \textbf{\bibinfo{volume}{A53}},
  \bibinfo{pages}{51} (\bibinfo{year}{1933}).

\bibitem[{\citenamefont{Tolman}(1934)}]{Tolman:1934}
\bibinfo{author}{\bibfnamefont{R.~C.} \bibnamefont{Tolman}},
  \bibinfo{journal}{Proc. Nat. Acad. Sci.} \textbf{\bibinfo{volume}{20}},
  \bibinfo{pages}{169} (\bibinfo{year}{1934}).

\bibitem[{\citenamefont{Bondi}(1947)}]{Bondi:1947}
\bibinfo{author}{\bibfnamefont{H.}~\bibnamefont{Bondi}}, \bibinfo{journal}{Mon.
  Not. Roy. Astron. Soc.} \textbf{\bibinfo{volume}{107}}, \bibinfo{pages}{410}
  (\bibinfo{year}{1947}).

\bibitem[{\citenamefont{Krasi\'nski}(1997)}]{Krasinski:1997}
\bibinfo{author}{\bibfnamefont{A.}~\bibnamefont{Krasi\'nski}},
  \emph{\bibinfo{title}{Inhomogeneous cosmological models}}
  (\bibinfo{publisher}{Cambridge University Press},
  \bibinfo{address}{Cambridge}, \bibinfo{year}{1997}).

\bibitem[{\citenamefont{Ellis et~al.}(1985)\citenamefont{Ellis, Nel, Maartens,
  Stoeger, and Whitman}}]{ellis:1985}
\bibinfo{author}{\bibfnamefont{G.~F.~R.} \bibnamefont{Ellis}},
  \bibinfo{author}{\bibfnamefont{S.~D.} \bibnamefont{Nel}},
  \bibinfo{author}{\bibfnamefont{R.}~\bibnamefont{Maartens}},
  \bibinfo{author}{\bibfnamefont{W.~R.} \bibnamefont{Stoeger}},
  \bibnamefont{and} \bibinfo{author}{\bibfnamefont{A.~P.}
  \bibnamefont{Whitman}}, \bibinfo{journal}{Phys. Rep.}
  \textbf{\bibinfo{volume}{124}}, \bibinfo{pages}{315} (\bibinfo{year}{1985}).

\bibitem[{\citenamefont{Humphreys et~al.}(1997)\citenamefont{Humphreys,
  Maartens, and Matravers}}]{Humphreys:1996fd}
\bibinfo{author}{\bibfnamefont{N.~P.} \bibnamefont{Humphreys}},
  \bibinfo{author}{\bibfnamefont{R.}~\bibnamefont{Maartens}}, \bibnamefont{and}
  \bibinfo{author}{\bibfnamefont{D.~R.} \bibnamefont{Matravers}},
  \bibinfo{journal}{Astrophys. J.} \textbf{\bibinfo{volume}{477}},
  \bibinfo{pages}{47} (\bibinfo{year}{1997}), \eprint{astro-ph/9602033}.

\bibitem[{\citenamefont{Ellis}(1971)}]{Ellis71}
\bibinfo{author}{\bibfnamefont{G.~F.~R.} \bibnamefont{Ellis}}, in
  \emph{\bibinfo{booktitle}{General relativity and cosmology: Proceedings of
  the international school of physics Enrico Fermi, course XLVII}}, edited by
  \bibinfo{editor}{\bibfnamefont{B.~K.} \bibnamefont{Sachs}}
  (\bibinfo{publisher}{Academic Press}, \bibinfo{year}{1971}), pp.
  \bibinfo{pages}{104--182}.

\bibitem[{\citenamefont{{Weinberg}}(1972)}]{Weinberg:1972}
\bibinfo{author}{\bibfnamefont{S.}~\bibnamefont{{Weinberg}}},
  \emph{\bibinfo{title}{{Gravitation and Cosmology: Principles and Applications
  of the General Theory of Relativity}}} (\bibinfo{publisher}{John Wiley \&
  Sons}, \bibinfo{year}{1972}).

\bibitem[{\citenamefont{Bennett et~al.}(1996)}]{Bennett:1996ce}
\bibinfo{author}{\bibfnamefont{C.~L.} \bibnamefont{Bennett}}
  \bibnamefont{et~al.}, \bibinfo{journal}{Astrophys. J.}
  \textbf{\bibinfo{volume}{464}}, \bibinfo{pages}{L1} (\bibinfo{year}{1996}),
  \eprint{astro-ph/9601067}.

\end{thebibliography}
   
\end{document}